\documentclass[prd,aps,showpacs,nofootinbib,twocolumn]{revtex4}%
\usepackage{graphicx}
\usepackage{amsmath}
\usepackage{amsfonts}
\usepackage{amssymb}
\usepackage{float}
\usepackage{color}%
\providecommand{\U}[1]{\protect\rule{.1in}{.1in}}
\providecommand{\U}[1]{\protect\rule{.1in}{.1in}}
\begin{document}
\title{Relativistic motion under constant force: velocity and acceleration behavior}
\author{Jhonnatan G. Pereira, Victor E. Mouchrek-Santos, Manoel M. Ferreira Jr}
\affiliation{Departamento de F\'{\i}sica, Universidade Federal do Maranh\~{a}o (UFMA),
Campus Universit\'{a}rio do Bacanga, S\~{a}o Lu\'{\i}s-MA, 65085-580 - Brasil}

\begin{abstract}
In relativistic dynamics, force and acceleration are no longer parallel. In
this article, we revisit the relativistic motion of a particle under the
action of a constant force, $\boldsymbol{f}$. \ For a two-dimensional motion,
the final velocity in each axis is $(f_{i}/f)c,$ independently of the initial
velocities, yielding an asymptotic velocity always parallel to the force. The
particular case in which the force is applied in a single axis is analyzed in
detail, with the behavior of velocity and acceleration being exhibited for
several configurations. Some previous results of the literature concerning
velocity and acceleration behavior are improved and better explored.
Differently from which was previously claimed, it is shown that a negative
acceleration component can exist in the direction of the biggest force
component and that acceleration does not decrease monotonically to zero.

\end{abstract}
\email{jhonnatangomes05@gmail.com, victor\_mouchrek@hotmail.com, manojr.ufma@gmail.com}
\maketitle

\section{Introduction}

In relativistic dynamics \cite{Rindler},\cite{Aharoni},\cite{French}%
,\cite{Hagedorn}, the usual concept of mass needs to be changed in order to
preserve conservation of linear momentum, in such a way it becomes a function
of velocity, $m(\mathrm{v})=m_{0}\gamma(\mathrm{v}),$ with $\gamma
(\mathrm{v})=1/\sqrt{1-\mathbf{v}^{2}/c^{2}}$. The 3-momentum and energy are
accordingly redefined, $\boldsymbol{p}=m(\mathrm{v})\mathbf{v}$,
$E=m_{0}\gamma(\mathrm{v})c^{2},$ with the 3-force given as $\boldsymbol{f}%
=d\boldsymbol{p}/dt$, or $\ \boldsymbol{f}=m\boldsymbol{a}+\dot{m}\mathbf{v}.$
Relativistic dynamics has been studied in several interesting respects,
involving methods for deriving force and momentum transformations
\cite{Martins},\cite{Stewart},\cite{Momentum}, relativistic mass \cite{Mass},
and the description of relativistic collisions \cite{Collision1},
\cite{Collision2}.

The mass dependence on the velocity yields $\dot{m}=m_{0}\dot{\gamma}%
=m\gamma^{2}(\mathbf{v}\cdot\boldsymbol{a})\boldsymbol{/}\boldsymbol{c}^{2}$
($\dot{m}=dm/dt)$, so that the relativistic 3-force presents two components,
\begin{equation}
\boldsymbol{f}=m\boldsymbol{a}+m\gamma^{2}(\mathbf{v}\cdot\boldsymbol{a}%
)\mathbf{v}\boldsymbol{/}c^{2}, \label{eq:3force}%
\end{equation}
one in the direction of acceleration, other parallel to velocity, becoming the
force, in general, no longer parallel to the acceleration engendered by
itself. This is a relativistic feature with no analogue in Newtonian dynamics,
opening a diversity of dynamic scenarios very different from the usual
ones.\ Taking the internal product of the force $\boldsymbol{f}$ with
$\mathbf{v},$ in eq. (\ref{eq:3force}), it results $(\boldsymbol{f}%
\cdot\mathbf{v)}=m\gamma^{2}(\mathbf{v}\cdot\boldsymbol{a})$. Thus, eq.
(\ref{eq:3force}) can be rewritten as:
\begin{equation}
\boldsymbol{f}=m\boldsymbol{a}+(\boldsymbol{f}\cdot\mathbf{v})\mathbf{v}%
\boldsymbol{/}c^{2}. \label{eq:3forcepower}%
\end{equation}
Taking the internal product of eq. (\ref{eq:3forcepower}) with force, it
implies $\boldsymbol{f}^{2}-m(\boldsymbol{a\cdot f)}=(\boldsymbol{f}%
\cdot\mathbf{v})^{2}\boldsymbol{/}c^{2}.$ Since \linebreak$(\boldsymbol{f}%
\cdot\mathbf{v})^{2}\boldsymbol{/}c^{2}\leq\boldsymbol{f}^{2},$ one has
$(\boldsymbol{a\cdot f)}>0,$ revealing that $\boldsymbol{a}$ and
$\boldsymbol{f}$ define an acute angle, as demonstrated in Ref. \cite{Redding}%
. Another contribution investigated the orientation of the acceleration
components relative to the force and its dependence on the system of reference
\cite{Lock}.

One of the first studies examining the intricate relation between force and
acceleration in relativistic dynamics is due to Tolman \cite{Tolman}, which
investigated the condition to have acceleration only in one axis. Tolman
noticed that applying force in one direction causes acceleration in this axis
and mass increases. Since the 3-momentum is conserved in the direction where
there is no applied force, the velocity in that direction should decrease in
order to assure a constant momentum. The x and y components of eq.
(\ref{eq:3force}) are%
\begin{align}
f_{x}  &  =ma_{x}+m\gamma^{2}(\mathrm{v}_{x}a_{x}+\mathrm{v}_{y}%
a_{y})\mathrm{v}_{x}\boldsymbol{/}c^{2},\\
f_{y}  &  =ma_{y}+m\gamma^{2}(\mathrm{v}_{x}a_{x}+\mathrm{v}_{y}%
a_{y})\mathrm{v}_{y}\boldsymbol{/}c^{2}.
\end{align}
Supposing an applied force in two dimensions, $\boldsymbol{f}%
=(f\boldsymbol{_{x}},f\boldsymbol{_{y}}),$ for acceleration only in the
y-axis, $a_{x}=0,$ Tolman wrote the ratio,%
\begin{equation}
\frac{f_{x}}{f_{y}}=\frac{\gamma^{2}\mathrm{v}_{y}\mathrm{v}_{x}%
\boldsymbol{/}c^{2}}{1+\gamma^{2}\mathrm{v}_{y}^{2}\boldsymbol{/}c^{2}}%
=\frac{\mathrm{v}_{y}\mathrm{v}_{x}}{c^{2}-\mathrm{v}_{x}^{2}},
\end{equation}
which states the relation to be fulfilled to guarantee $a_{x}=0,$ involving
force applied in both axis. Tolman also observed \cite{Tolman2} the existence
of two situations where force and acceleration remain parallel: (i) when force
is applied longitudinally to the direction of the motion ($\boldsymbol{f}%
\parallel\mathbf{v}$), (ii) when force is applied in a direction orthogonal to
the velocity of the particle, making it describe a circular motion
($\boldsymbol{f}\bot\mathbf{v}$). In both situations, the initial geometrical
configuration is maintained in time: the force is kept parallel to
acceleration as the motion evolves. For force parallel to velocity, eq.
(\ref{eq:3forcepower}) provides $\boldsymbol{f}=m_{\parallel}\boldsymbol{a}$,
with $m_{\parallel}=m_{0}\gamma^{3}$ being the longitudinal mass; for force
orthogonal to velocity, it holds $\boldsymbol{f}=m_{\bot}a$, where $m_{\bot
}=m_{0}\gamma$ is orthogonal mass, with ($m_{\bot}<m_{\parallel}$).

In most relativistic scenarios, force and velocity are not parallel. The
richness of these situations was already addressed in some respects
\cite{Redding}. One of the remarkable consequences is the existence of a
negative acceleration component \cite{Lock},\cite{Ficken}, in two-dimensional
systems with force applied in both orthogonal axis, $f_{x}>0,$ $f_{y}>0.$ This
motion was examined by plotting the value of the ratio $a_{y}/a_{x\text{ }}%
$for several values of\ velocity ($\mathrm{v}_{x\text{ }}$and $\mathrm{v}%
_{y})$ in the interval $\left[  0,\sqrt{c^{2}/2}\right]  ,$ for different
values of the ratio $R=f_{x}/f_{y}$ \cite{Ficken}. A similar analysis was
developed in three dimensions \cite{Gonzalez}, so that the ratios $a_{x}%
/a_{y}$ and $a_{z}/a_{y}$ were plotted for values of velocities $\mathrm{v}%
_{x}$, $\mathrm{v}_{y}$ and $\mathrm{v}_{z}$ in the interval $[0,\sqrt
{c^{2}/3}]$, for different values of the ratios $R_{1}=f_{x}/f_{y}$,
$R_{2}=f_{y}/f_{z}$. These authors asserted that the negative acceleration
appears in the direction where the applied force is smaller, which is not
always true, however. This kind of analysis, based on the ratio $R=f_{x}%
/f_{y},$ is restricted in some respects, namely: (i) the components of
velocity are made equal ($\mathrm{v}_{x}=\mathrm{v}_{y})$ all the time in
order to simplify the analysis, which exclude interesting situations where the
velocities differ from each other; (ii) the behavior of the velocity
components is not properly discussed; (iii) by studying the ratio between
accelerations, it is not clear what happens with each acceleration component.

In this work, we reassess the description of a two dimensional relativistic
motion under the action of a constant force, combining two different
approaches, an analytical and numerical procedure, in order to shed light on
the three main aspects highlighted above, which remain barely discussed until
this moment.

\section{Two-dimensional Motion of a particle under the action of a
relativistic 3-force}

In this section, we develop an analysis that goes beyond the aspects
scrutinized in the previous works about force and acceleration in special
relativity \cite{Ficken}, \cite{Gonzalez}.\ Remembering that $\ \boldsymbol{f}%
=(\boldsymbol{f}_{x},\boldsymbol{f}_{y})=(d\boldsymbol{p}_{x}%
/dt,d\boldsymbol{p}_{y}/dt),$ and regarding the case the force is constant,
one achieves%

\begin{align}
p_{x}(t)  &  =p_{0x}+f_{x}t,\label{eq:xmomentum}\\
p_{y}(t)  &  =p_{0y}+f_{y}t, \label{eq:ymomentum}%
\end{align}
in which $p_{0x}$ and $p_{0y}$ are the initial x and y components of momentum,
respectively. Knowing that the particle velocity is $\mathbf{v}=c^{2}%
\mathbf{p/E,}$ where $\mathbf{p}$ is the 3-momentum and $E=\sqrt{m_{0}%
^{2}c^{4}+\mathbf{p}^{2}c^{2}}$ is the relativistic energy, with
$\mathbf{p}^{2}=p_{x}^{2}+p_{y}^{2}$, the velocity components are given as%

\begin{equation}
\mathrm{v}_{i}(t)=\frac{(p_{0i}+f_{i}t)c^{2}}{\sqrt{m_{0}^{2}c^{4}%
+\mathbf{p}^{2}c^{2}}}, \label{eq:vx}%
\end{equation}
where $i=1,2$ for $x$ and $y$ components. The direction of the force is given
by the angle $\theta$ it forms with the $x$-axis, where $\tan\theta
=f_{y}/f_{x}$. In the same way, the velocity of the particle points in the
direction of the $\alpha$ angle, written as%

\begin{equation}
\tan\alpha=\frac{\mathrm{v}_{y}}{\mathrm{v}_{x}}=\frac{p_{0y}+f_{y}t}%
{p_{0x}+f_{x}t}. \label{eq:velocityratio}%
\end{equation}
In the asymptotic limit ($t\rightarrow\infty)$, \ eq. (\ref{eq:velocityratio})
tends to $f_{y}/f_{x}$, therefore $\alpha\rightarrow\theta$: the direction of
the velocity approaches asymptotically the direction of the applied force. It
is thus demonstrated that the velocity will always tend to be parallel to
force after a long enough time. If the same limit $t\rightarrow\infty$ is
applied to the velocity components (\ref{eq:vx}), we attain the following
asymptotic values:%

\begin{equation}
\mathrm{v}_{x}(\infty)=\frac{f_{x}c}{f},\text{ }\mathrm{v}_{y}(\infty
)=\frac{f_{y}c}{f}, \label{eq:vxlimit}%
\end{equation}
where $f=\sqrt{f_{x}^{2}+f_{y}^{2}}$ is the 3-force magnitude. Eq.
(\ref{eq:vxlimit}) shows that the final value of the velocity is always $c$,
that is, $\mathrm{v}(\infty)=\sqrt{\mathrm{v}_{x}^{2}(\infty)+\mathrm{v}%
_{y}^{2}(\infty)}=c.$ Further, the final velocity in each axis, $(f_{i}/f)c,$
depends on the value of the corresponding component of force, independently of
the initial velocities. If there is force applied only in one axis, the final
velocity in this axis will be equal to $c$, being null in the other axis. The
cases where there is force in only one axis are well appropriate to
demonstrate the manifestation of the negative acceleration component, and will
be discussed in the next sections.

\subsection{Velocity behavior}

A particle of rest mass $m_{0},$ with \ initial velocity $\mathbf{v}%
_{0}=(\mathrm{v}_{0x},\mathrm{v}_{0y})$, when accelerated under the action of
a\textbf{ }constant 3-force $f=(f_{x},f_{y}),$ reaches asymptotic velocity
components equal to $(f_{i}/f)c$\, with $f_{i}$ being the force component in
the $i$-axis. We know the initial and final velocities, but some aspects of
the motion evolution need to be better elucidated and discussed. As observed
by Tolman \cite{Tolman},\cite{Tolman2}, if a force is applied in a single
direction, there will appear a (negative) acceleration in the orthogonal axis.
The situation ($f_{x}>0,f_{y}=0)$, corresponding to $R=\infty,$ was in
principle depicted in the graph of Ref. \cite{Ficken}, showing the existence
of negative acceleration in the y-axis. No additional details were explored,
though, and are the goal of the present work.

Consider Eq. (\ref{eq:3forcepower}), written for the acceleration components:%

\begin{align}
a_{x}  &  =\frac{1}{m}\left[  f_{x}(1-\mathrm{v}_{x}^{2}/c^{2})-f_{y}%
\mathrm{v}_{x}\mathrm{v}_{y}/c^{2}\right]  ,\label{eq:ax}\\
a_{y}  &  =\frac{1}{m}\left[  f_{y}(1-\mathrm{v}_{y}^{2}/c^{2})-f_{x}%
\mathrm{v}_{x}\mathrm{v}_{y}/c^{2}\right]  . \label{eq:ay}%
\end{align}
Now, we specify the case where force is applied only along the y-axis,\textbf{
}$\boldsymbol{f}=(0,f\boldsymbol{_{y}}),$ with $\boldsymbol{f_{y}} > 0$,
because this choice allows us to notice clearly the effects of negative
acceleration. Eq. (\ref{eq:vxlimit}) shows that the particle will be
progressively accelerated in the y-axis, having asymptotic velocity
$\mathrm{v}_{y}(\infty)=c$, while its component $\mathrm{v}_{x\text{ }}%
$decreases, tending to zero. For this situation Eqs. (\ref{eq:ax}) and
(\ref{eq:ay}) simplify to:%

\begin{align}
\frac{d\mathrm{v}_{x}}{dt}  &  =-\frac{f_{y}\mathrm{v}_{x}\mathrm{v}_{y}%
}{mc^{2}},\label{eq:ax_simple}\\
\frac{d\mathrm{v}_{y}}{dt}  &  =\frac{f_{y}(1-\mathrm{v}_{y}^{2}/c^{2})}{m},
\label{eq:ay_simple}%
\end{align}
wherein it was replaced $a_{x}=d\mathrm{v}_{x}/dt,$ $a_{y}=d\mathrm{v}_{y}%
/dt$. These equations will be used to reveal some aspects not yet explored of
this two-dimensional motion.

\begin{figure}[h]
\centering\includegraphics[scale=0.3]{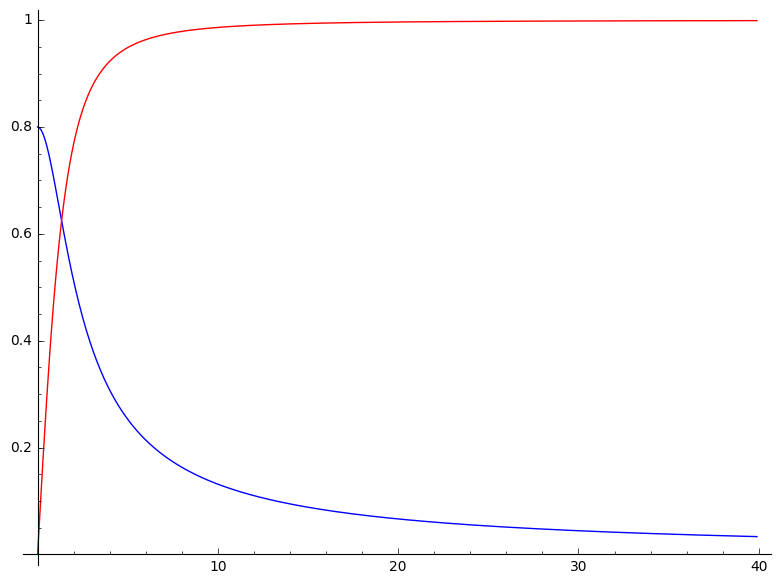} \caption{Graph of velocities
$\mathrm{v}_{x}$, $\mathrm{v}_{y}$ $\times$ time. The red and blue line
represent $\mathrm{v}_{y}$ and $\mathrm{v}_{x}$, respectively. The initial
velocities are $\mathrm{v}_{x}(0)=0.8c$ and $\mathrm{v}_{y}(0)=0$.}%
\label{fig:case1}%
\end{figure}

From the analysis of eqs. (\ref{eq:ax_simple}) and (\ref{eq:ay_simple}), we
see that the y-acceleration is always positive regardless \textcolor{red}{of}
the velocity. The x-acceleration, however, behaves differently, depending on
the product $\mathrm{v}_{x}\mathrm{v}_{y}$. If $\mathrm{v}_{x}(0)>0$ and
$\mathrm{v}_{y}(0)=0$ the component $a_{x}$ is initially null and negative
afterwards. This makes the velocity $\mathrm{v}_{x}$ to be initially constant
and decreasing later\textbf{.} This initial behavior of velocity is
demonstrated by the little plateau at $t=0$, see Fig. \ref{fig:case1}.

\begin{figure}[h]
\centering\includegraphics[scale=0.3]{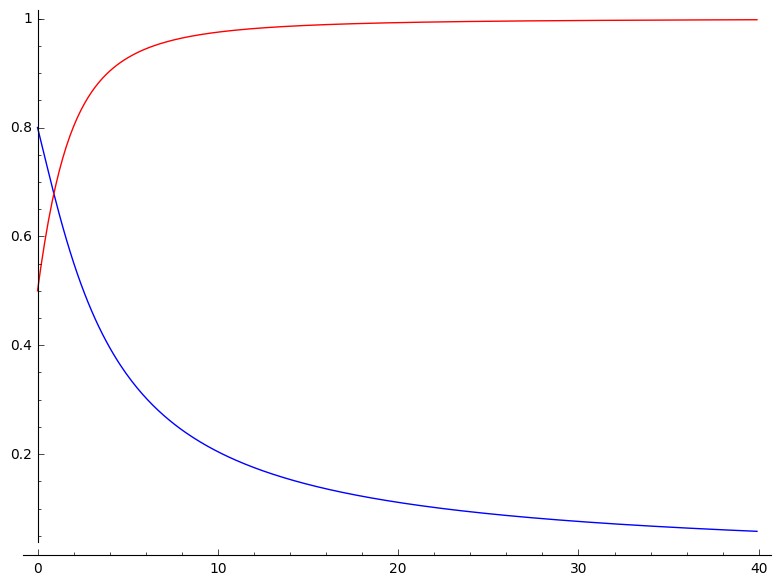} \caption{Graph of velocities
$\mathrm{v}_{x}$, $\mathrm{v}_{y}\times$ time. The red and blue line represent
$\mathrm{v}_{y}$ and $\mathrm{v}_{x}$, respectively. The initial velocities
are $\mathrm{v}_{x}(0)=0.8c$ and $\mathrm{v}_{y}(0)=0.5c$.}%
\label{fig:case1b}%
\end{figure}

\begin{figure}[h]
\centering\includegraphics[scale=0.3]{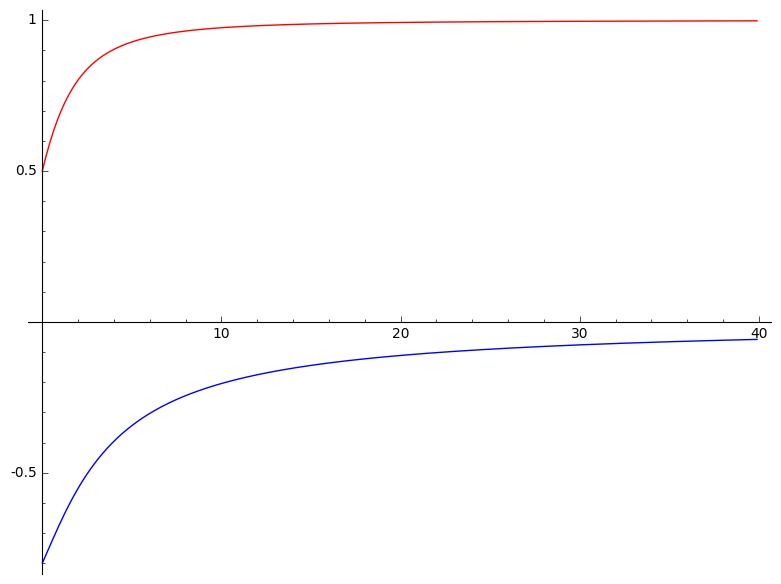} \caption{Graph of
velocities $\mathrm{v}_{x}$, $\mathrm{v}_{y}$ $\times$ time. The red and blue
line represent $\mathrm{v}_{y}$ and $\mathrm{v}_{x}$, respectively. The
initial velocities are $\mathrm{v}_{x}(0)=-0.8c$ and $\mathrm{v}_{y}%
(0)=0.5c$.}%
\label{fig:case4}%
\end{figure}

If $\mathrm{v}_{x}(0)>0$ and $\mathrm{v}_{y}(0)>0$, the only difference in
relation to the first case is that $\mathrm{v}_{x}$ decreases immediately from
the start, as displayed in Fig. \ref{fig:case1b}. When $\mathrm{v}_{x}(0)<0$,
$\mathrm{v}_{y}(0)>0$, Eq. (\ref{eq:ax_simple}) shows that $a_{x}$ is
positive. Since $\mathrm{v}_{x}$ is negative, the particle will decelerate in
the x-axis, tending continuously to zero, as shown in Fig. \ref{fig:case4}.

\begin{figure}[h]
\centering\includegraphics[scale=0.3]{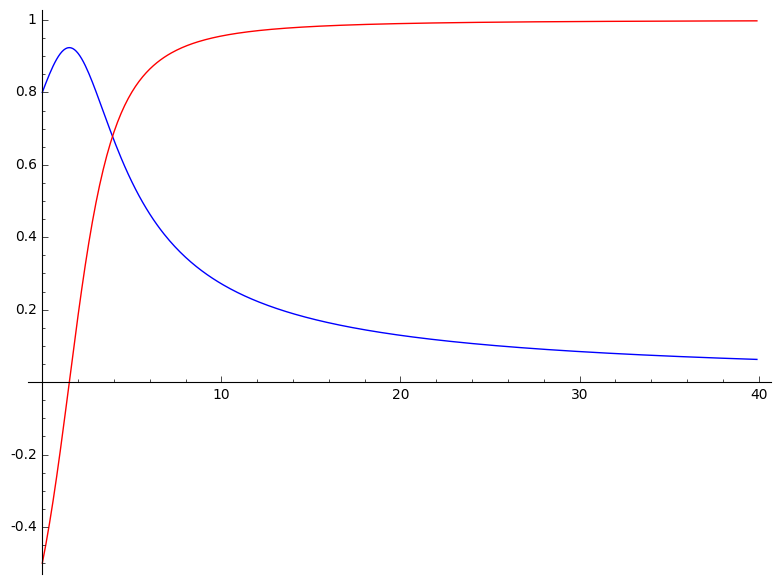}\caption{Graph of velocities
$\mathrm{v}_{x}$, $\mathrm{v}_{y}$ x time. The red line represents
$\mathrm{v}_{y}$, while the blue one represents $\mathrm{v}_{x}$. The initial
velocities are $\mathrm{v}_{x}(0)=0.8c$ and $\mathrm{v}_{y}(0)=-0.5c$.}%
\label{fig:case7}%
\end{figure}

\begin{figure}[h]
\centering\includegraphics[scale=0.3]{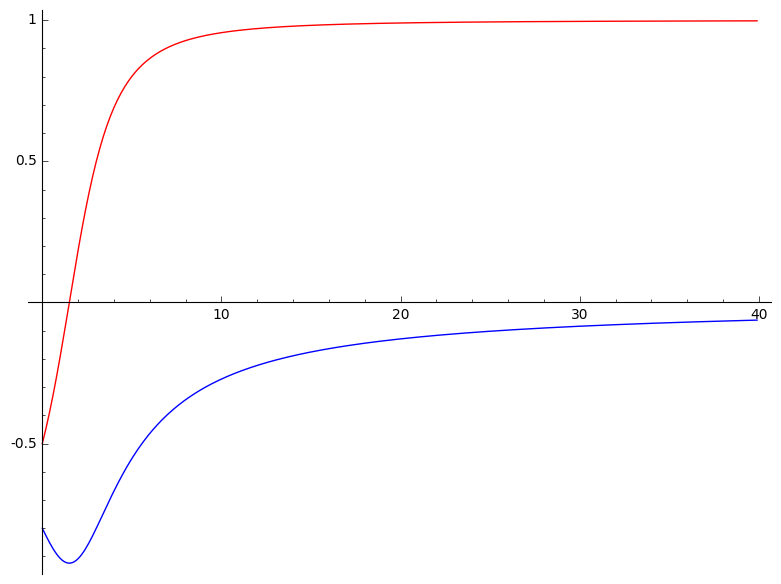} \caption{Graph of velocities
$\mathrm{v}_{x}$, $\mathrm{v}_{y}$ x time. The red and blue line represent
$\mathrm{v}_{y}$ and $\mathrm{v}_{x}$, respectively. The initial velocities
are $\mathrm{v}_{x}(0)=-0.8c$ and $\mathrm{v}_{y}(0)=-0.5c$.}%
\label{fig:case3}%
\end{figure}

\begin{figure}[h]
\centering\includegraphics[scale=0.3]{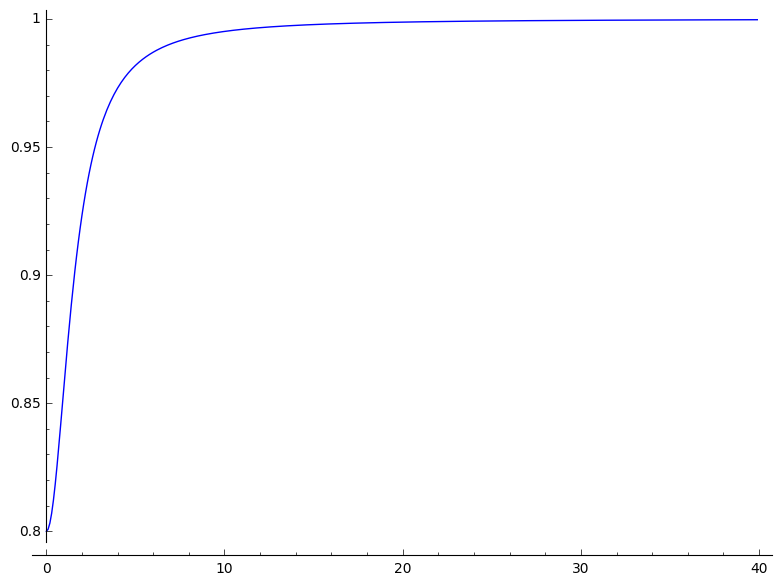}\caption{Graph of $|\mathbf{v}|$
against time. This graph corresponds to the situation depicted in Fig.
\ref{fig:case1}.}%
\label{fig:av1}%
\end{figure}

\begin{figure}[h]
\centering\includegraphics[scale=0.3]{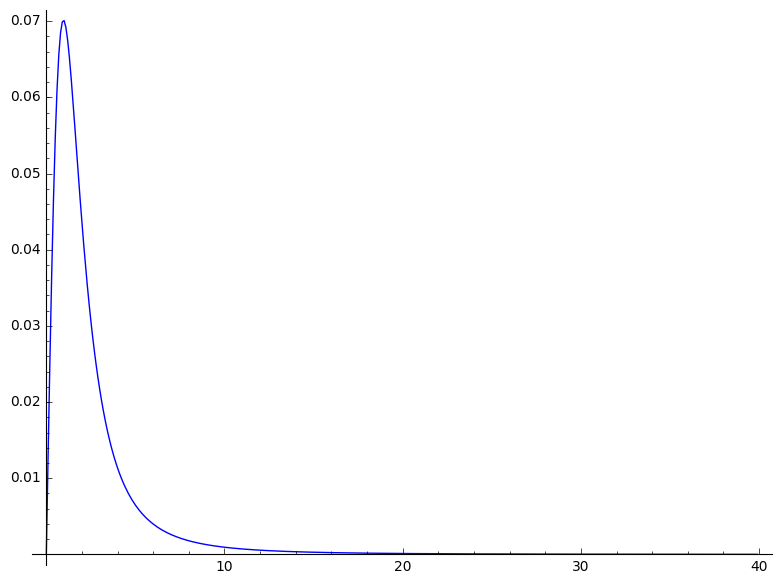} \caption{Graph of
$(\boldsymbol{a}\cdot\mathbf{v})$ against time. This graph corresponds to the
situation depicted in Fig. \ref{fig:case1}.}%
\label{fig:av1a}%
\end{figure}

Another interesting situation is the one where $\mathrm{v}_{y}(0)<0$ and
$\mathrm{v}_{x}(0)>0$. In that case, we see from eq. (\ref{eq:ax_simple}) that
$a_{x}$ is initially positive and therefore the particle initially accelerates
in the x-axis, as shown in Fig. \ref{fig:case7}. With the force acting on the
positive sense of y-axis, it will reverse the direction of motion in the
y-axis: $\mathrm{v}_{y}$ becomes positive and the particle starts to
decelerate in the x-axis, similarly to the other cases previously discussed.
For the case where both initial velocities are negative, $\mathrm{v}_{y}(0)<0$
and $\mathrm{v}_{x}(0)<0$, $a_{x}$ is initially negative. The particle will
first accelerate in x and then decelerate, but with $\mathrm{v}_{x}$ being
negative, as illustrated in Fig. \ref{fig:case3}. These last two cases exhibit
an initial acceleration regime followed by a deceleration one, behavior that
was not reported in the literature yet.

\begin{figure}[h]
\centering\includegraphics[scale=0.3]{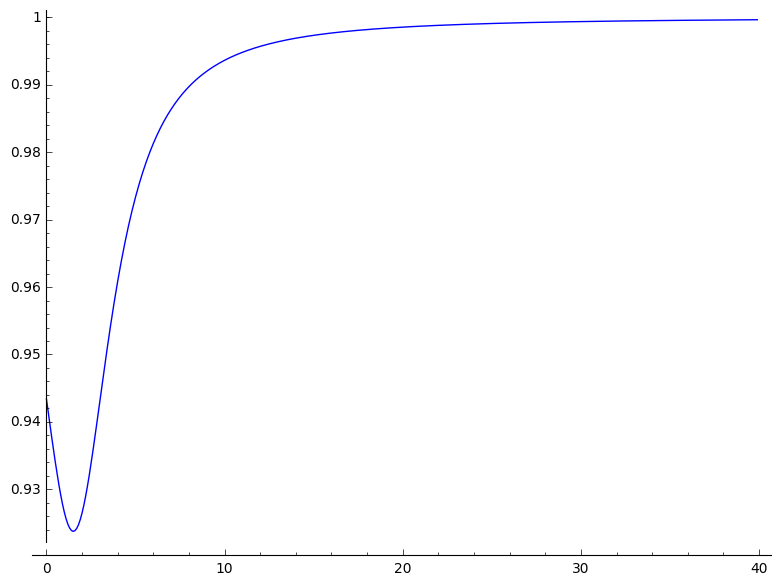} \caption{Graph of
$|\mathbf{v}|$ against time. This graph corresponds to the situation depicted
in Figs. \ref{fig:case7} and \ref{fig:case3}.}%
\label{fig:av2}%
\end{figure}

\begin{figure}[h]
\centering\includegraphics[scale=0.3]{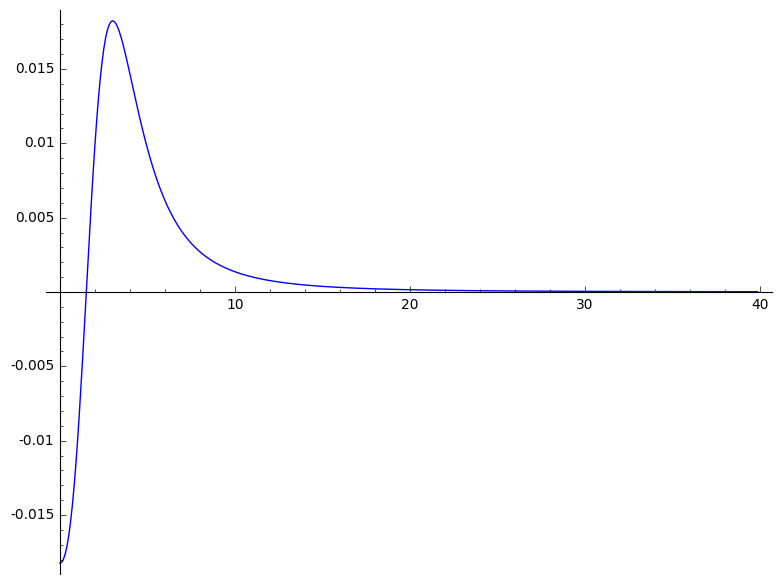} \caption{Graph of
$\boldsymbol{a}\cdot\mathbf{v}$ against time. This graph corresponds to the
situation depicted in Figs. \ref{fig:case7} and \ref{fig:case3}.}%
\label{fig:av2a}%
\end{figure}

As shown by Redding \cite{Redding}, the expression $m_{0}\gamma^{3}%
(\mathbf{v}\cdot\mathbf{a})=(\boldsymbol{f}\cdot\mathbf{v})$ states the angle
between force and velocity is of the same type as the one between acceleration
and velocity. By analyzing the situation depicted in Fig. \ref{fig:case1}, we
see that acceleration and velocity are initially perpendicular. After the
initial moment, the particle presents a non null $\mathrm{v}_{y}$ and the
angle between force and velocity becomes acute. The same holds for the angle
between acceleration and velocity, in such a way that $(\mathbf{v}%
\cdot\mathbf{a})>0,$ which implies increasing $|\mathbf{v}|,$ since
$d(\mathbf{v}^{2})/dt=2(\mathbf{v}\cdot\mathbf{a}).$ This is the scenario
displayed in Figs. \ref{fig:av1} and \ref{fig:av1a}, which shows the behavior
of $(\mathbf{v}\cdot\mathbf{a})$ and $|\mathbf{v}|$ for the situation of Fig.
\ref{fig:case1}. When $\mathrm{v}_{y}(0)<0,$ force and velocity make initially
an obtuse angle, implying $(\mathbf{v}\cdot\mathbf{a})<0$, which yields a
decreasing $|\mathbf{v}|$. Such scenario can be seen in the initial part of
the graphs of Figs. \ref{fig:av2} and \ref{fig:av2a}, corresponding to the
situation represented in Figs. \ref{fig:case7} and \ref{fig:case3}. The
absolute value $|\mathbf{v}|$ decreases while\textbf{ }$(\mathbf{v}%
\cdot\mathbf{a})<0$, behavior that changes when $\mathrm{v}_{y}=0$. After
$\mathrm{v}_{y}$ changes sign, becoming positive, one has $(\boldsymbol{f}%
\cdot\mathbf{v})>0$ and $(\mathbf{v}\cdot\mathbf{a})>0,$ so that
$|\mathbf{v}|$ increases continuously to its asymptotic value.

\subsection{Acceleration behavior}

In Ref. \cite{Redding}, it is claimed that the acceleration $\boldsymbol{a}$
is a decreasing function of velocity, tending monotonically to zero when
$\beta=\mathrm{v}/c$ tends to unity. More specifically, we can say that
$\boldsymbol{a}$ magnitude goes to zero when the velocity components tend to
their respective asymptotic values, but it does not go in a monotonic way.
This fact, already highlighted in the last section, was not noticed in Ref.
\cite{Redding} and will be explored in this section. Taking the scalar product
of Eq. \ref{eq:3force} with the acceleration,%

\begin{equation}
\boldsymbol{f}\cdot\boldsymbol{a}=m\boldsymbol{a}^{2}+m\gamma^{2}%
(\mathbf{v}\cdot\boldsymbol{a})^{2}\boldsymbol{/}c^{2},
\end{equation}
and $(\boldsymbol{f}\cdot\mathbf{a}\boldsymbol{)}=\left\vert \boldsymbol{f}%
\right\vert \left\vert \mathbf{a}\right\vert \cos\phi,$ $(\mathbf{v}%
\cdot\mathbf{a})=\left\vert \mathbf{v}\right\vert \left\vert \mathbf{a}%
\right\vert \cos\sigma,$ the resulting magnitude of the acceleration is
\begin{equation}
\left\vert \boldsymbol{a}\right\vert =\frac{\left\vert \boldsymbol{f}%
\right\vert \cos\phi}{m[1+\gamma^{2}\left\vert \mathbf{v}\right\vert ^{2}%
\cos\sigma\boldsymbol{/}c^{2}]},\text{ } \label{FR15B}%
\end{equation}
which depends on $\left\vert \mathbf{v}\right\vert ^{2}$ and on the factors
$\cos\phi$ and $\cos\sigma,$ that also change with time. As\textbf{
}$\left\vert \mathbf{v}\right\vert ^{2}$\textbf{ }does not have a monotonic
behavior, the same can be expected for\textbf{ }$\left\vert \boldsymbol{a}%
\right\vert .$\textbf{ }Further, Ficken \cite{Ficken} and Gonzalez
\cite{Gonzalez} claimed that under the action of a relativistic 3-force, there
will exist a negative acceleration in the direction of the smallest force
component, which is not exactly correct. To investigate the conditions in
which the acceleration becomes negative, we differentiate the velocity
components (\ref{eq:vx}), obtaining:%

\begin{align}
a_{x}  &  =\frac{c}{(m_{0}^{2}c^{2}+\mathbf{p}^{2})^{3/2}}[\alpha f_{y}%
t+f_{x}m_{0}^{2}c^{2}+\alpha p_{0y}],\label{eq:analyticalax}\\
a_{y}  &  =\frac{c}{(m_{0}^{2}c^{2}+\mathbf{p}^{2})^{3/2}}[-\alpha
f_{x}t+f_{y}m_{0}^{2}c^{2}-\alpha p_{0x}], \label{eq:analyticalay}%
\end{align}
with
\begin{equation}
\alpha=\left(  f_{x}p_{0y}-f_{y}p_{0x}\right)  .
\end{equation}

Eqs. (\ref{eq:analyticalax}) and (\ref{eq:analyticalay}) reveal that
the\textbf{ }appearance of negative acceleration is not determined only by the
force magnitude or the ratio $R=f_{x}/f_{y},$ as claimed in refs.
\cite{Gonzalez} and \cite{Ficken}, but by a combination of factors involving
the initial momentum components as well. In the general two-dimensional case,
$\boldsymbol{f}=(f_{x},f_{y}),$ with $f_{x}\geq0,f_{y}\geq0$, the factor
$\alpha$ can take on both signs.

\subsubsection{Force in two directions}

\textbf{Case 1}: we begin analyzing the situation $\alpha>0$ or $R>p_{0x}%
/p_{0y}$, that is,%
\begin{equation}
\text{\ }\frac{f_{x}}{f_{y}}>\frac{p_{0x}}{p_{0y}}, \label{cond1}%
\end{equation}
for which there are two main possibilities.

Situation (a): for $p_{0y}>0,$ we have $a_{x}>0$ for any time ($t\geq0$). In
the y-axis, $a_{y}$ can begin positive or negative, depending on the sign of
the factor $(f_{y}m_{0}^{2}c^{2}-\alpha p_{0x}).$ If $a_{y}(0)>0$, this
acceleration component will become negative after $t=(f_{y}m_{0}^{2}%
c^{2}-\alpha p_{0x})/\alpha f_{x}.$ Note that $p_{0x}<0$ and $p_{0y}>0$ also
fulfills condition (\ref{cond1}). One notices that one can have $a_{y}(t)<0$
even for $f_{y}>f_{x}$, thus constituting a scenario of negative acceleration
in the direction of the major force.

Situation (b): For $p_{0y}<0,$ we can have $a_{x}(0)<0,$ if $f_{x}m_{0}%
^{2}c^{2}<\alpha\left\vert p_{0y}\right\vert .$ In such a situation, the
acceleration $a_{x}$ will be negative until $t=-(f_{x}m_{0}^{2}c^{2}+\alpha
p_{0y})/\alpha f_{y},$becoming positive afterwards. Again, one can have
$p_{0x}<0$ or $p_{0x}>0$ in accordance with Eq. (\ref{cond1}). So $a_{y}$ can
begin positive or negative, respectively. For $a_{y}(0)>0$, this component
will become negative after the moment $t=(f_{y}m_{0}^{2}c^{2}-\alpha
p_{0x})/\alpha f_{x}.$ This takes place even for $f_{y}>f_{x}.\bigskip$

\textbf{Case 2}: The situation $\alpha<0$ or $R<p_{0x}/p_{0y}$, that is,%
\begin{equation}
\text{\ }\frac{f_{x}}{f_{y}}<\frac{p_{0x}}{p_{0y}},
\end{equation}
is similar to case 1, with $a_{x}$ and $a_{y}$ displaying the behavior of one
another. The component $p_{0x}$\textbf{ }determines whether $a_{y}$ is always
positive or starts negative and become positive in the sequel. Similarly,
$p_{0y}$ sign dictates if $a_{x}$ starts positive $\left(  p_{0y}<0\text{ or
}p_{0y}>0\text{ with }f_{x}m_{0}^{2}c^{2}>-\alpha p_{0y}\right)  $ and then
becomes negative or if $a_{x}$ is always negative$\left(  p_{0y}>0\text{ and
}f_{x}m_{0}^{2}c^{2}<\left\vert \alpha p_{0y}\right\vert \right)  $. The time
in which the acceleration components switch signs can be obtained by making
the substitutions $f_{x}\rightarrow f_{y}$ and $p_{0x}\rightarrow p_{0y}$ in
the previous equations of case 1.\bigskip

\textbf{Case 3}: An interesting situation happens for $\alpha=0,$ that is,
$R=p_{0x}/p_{0y}$, so that
\begin{align}
a_{x}  &  =\frac{c}{(m_{0}^{2}c^{2}+\mathbf{p}^{2})^{3/2}}[f_{x}m_{0}^{2}%
c^{2}],\\
a_{y}  &  =\frac{c}{(m_{0}^{2}c^{2}+\mathbf{p}^{2})^{3/2}}[f_{y}m_{0}^{2}%
c^{2}],
\end{align}
and no negative acceleration appears at all for any positive force components
( $f_{x}>0,f_{y}>0)$. This case is not included in previous analysis, as well.

\subsubsection{Force in only one direction}

In the circumstance where there is force applied in only one axis, the
acceleration becomes negative (not always) in the axis where there is no
force. For $\boldsymbol{f}=(0,f_{y})$, one has $\alpha=-f_{y}p_{0x},$ and
\begin{align}
a_{x}  &  =\frac{c\alpha}{(m_{0}^{2}c^{2}+\mathbf{p}^{2})^{3/2}}[f_{y}%
t+p_{0y}],\\
a_{y}  &  =\frac{c}{(m_{0}^{2}c^{2}+\mathbf{p}^{2})^{3/2}}[f_{y}m_{0}^{2}%
c^{2}-\alpha p_{0x}].
\end{align}
Concerning such a scenario, there are 4 configurations for the acceleration,
to be analysed as follows.\bigskip

\begin{figure}[h]
\centering\includegraphics[scale=0.3]{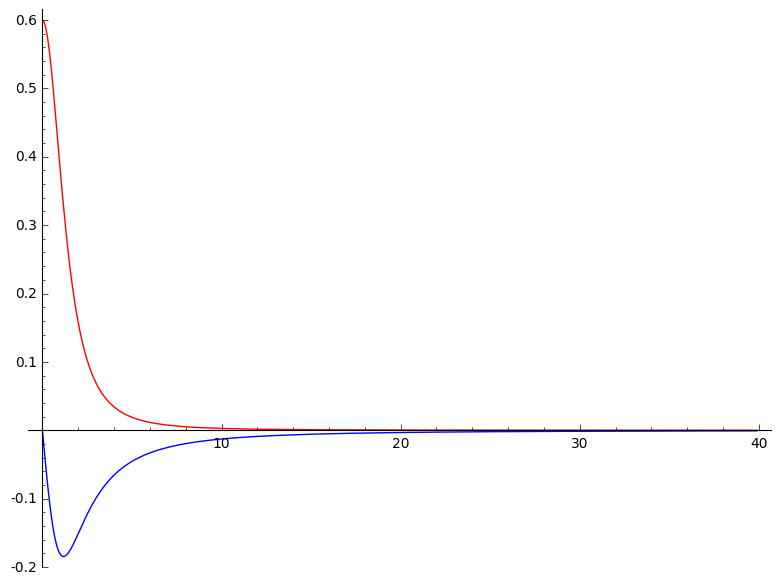}\caption{Graphs of $a_{x}$
$\times$ time (blue), $a_{y}$ $\times$ time (red), corresponding to the
scenario represented in Fig. \ref{fig:case1}; $\mathrm{v}_{x}(0)=0.8c,$
$\mathrm{v}_{y}(0)=0.$ }%
\label{fig:accel}%
\end{figure}

\textbf{Situation 1: }For $p_{0y}\geq0$ and $p_{0x}\geq0$ $\left(  \alpha
\leq0\right)  ,$ we notice that $a_{x}\leq0$ and $a_{y}(t)>0$ for any time.
This case is depicted in Fig. [\ref{fig:accel}], corresponding to the
situation of Fig. \ref{fig:case1}, with $\mathrm{v}_{x}(0)=0.8c,$
$\mathrm{v}_{y}(0)=0.$ Fig. \ref{fig:accel} shows clearly that the component
$a_{y}$ decreases monotonically, but the same does not happen for $a_{x}.$
Since $\mathrm{v}_{y}(0)=0$, from eq. (\ref{eq:ax_simple}) we have
$a_{x}(0)=0$.\textbf{ }After the movement begins,\textbf{ }\text{we have
}$\mathrm{v}$\text{$_{y}>0$, }$\mathrm{v}_{x}>0$, and Eq. (\ref{eq:ax_simple})
provides $a_{x}$ negative, that is, it starts from zero\textbf{,} decreases,
reaches a minimum value and then rises (absolute value diminishes), tending
continuously to zero. The component $a_{y}$, on the other hand, is always
positive, in accordance with Eq. (\ref{eq:ay_simple}). 

\bigskip\bigskip

\begin{figure}[h]
\centering\includegraphics[scale=0.3]{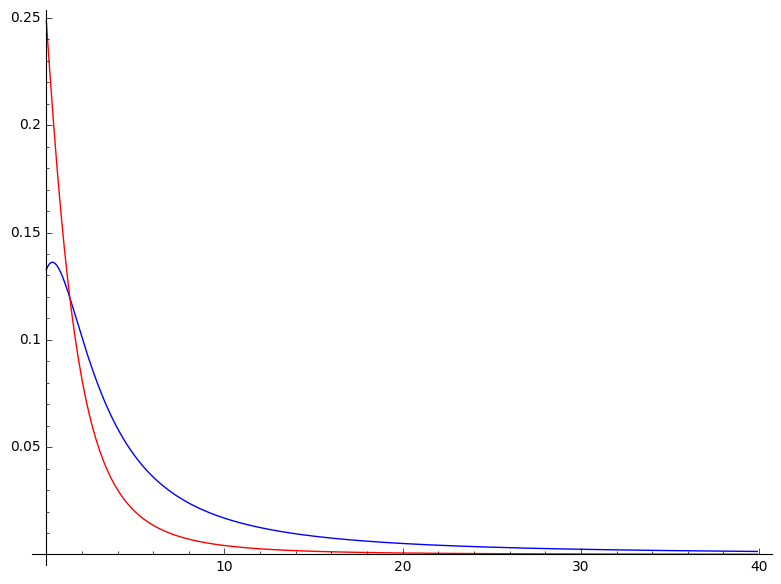}\caption{Graphs of $a_{x}$
$\times$ time (blue), $a_{y}$ $\times$ time (red), corresponding to the
scenario represented in Fig. \ref{fig:case4}; $\mathrm{v}_{x}(0)=-0.8c,$
$\mathrm{v}_{y}(0)=0.5c.$ }%
\label{fig:accel4}%
\end{figure}

\textbf{Situation 2: }For $p_{0y}>0$ and $p_{0x}<0$ $\left(  \alpha>0\right)
,$ we notice that $a_{x}(t)>0$ and $a_{y}(t)>0$ for any time. As the product
$\mathrm{v}_{x}\mathrm{v}_{y}$ starts negative, we see from eq.
(\ref{eq:ax_simple}) that $a_{x}(0)>0$. The signs of both acceleration
components are kept unaltered during the whole movement in this situation. In
this circumstance, despite the positive $a_{x}(t),$ the absolute value of
velocity $\mathrm{v}_{x}$ decreases steadily, due to the fact that
$\mathrm{v}_{x}$ is negative, and therefore the positive x-acceleration slows
down the particle in that direction. This scenario is depicted in the graph of
Fig. \ref{fig:case4} and the corresponding acceleration graph is shown in Fig.
\ref{fig:accel4}.

\bigskip

\begin{figure}[h]
\centering\includegraphics[scale=0.3]{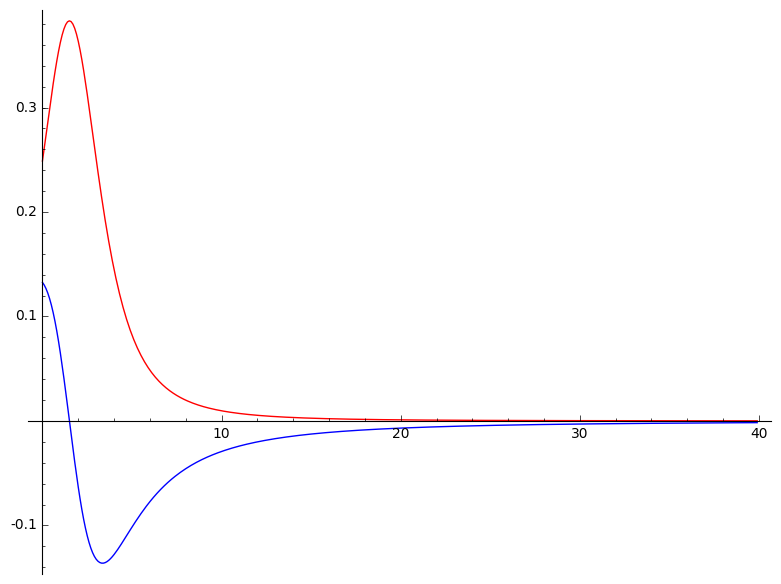} \caption{Graphs of $a_{x}$
$\times$ time (blue), $a_{y}$ $\times$ time (red), corresponding to the
scenario represented in Fig. \ref{fig:case7}; $\mathrm{v}_{x}(0)=0.8c,$
$\mathrm{v}_{y}(0)=-0.5c.$}%
\label{fig:accel2}%
\end{figure}

\textbf{Situation 3: }For $p_{0y}<0$ and $p_{0x}>0$ $\left(  \alpha<0\right)
,$ we notice that $a_{x}(0)>0$, but $a_{x}$ becomes negative from
$t=-p_{0y}/f_{y}$ on, while $a_{y}(t)>0$ for any time. This case is depicted
in Fig. [\ref{fig:accel2}], corresponding to the scenario of Fig.
\ref{fig:case7}, that is, $\mathrm{v}_{x}(0)=0.8c,$ $\mathrm{v}_{y}(0)=-0.5c.$
As the product $\mathrm{v}_{x}\mathrm{v}_{y}$ begins negative, eq.
(\ref{eq:ax_simple}) shows that x-acceleration is initially positive,
$a_{x}(0)>0$, being positive while $\mathrm{v}_{x}\mathrm{v}_{y}<0.$ As
$\mathrm{v}_{y}$ becomes less negative (increases) under the action of the
force, $a_{x}$ decreases, becoming null when $\mathrm{v}_{y}=0.$ From Eq.
(\ref{eq:ay_simple}), one notices that the smaller $\mathrm{v}_{y}^{2}$ is,
the bigger is the factor $(1-\mathrm{v}_{y}^{2}/c^{2})$ and the component
$a_{y}$. For this reason, $a_{y}$ reaches a maximum when $\mathrm{v}_{y}=0$,
at the same point in which $a_{x}=0$. From this point on the product
$\mathrm{v}_{x}\mathrm{v}_{y}$ is positive, so that the acceleration profiles
recover the same behavior of Fig. \ref{fig:accel}: the component $a_{y}$
diminishes monotonically; $a_{x}$ decreases, reaches a minimum value and then
rises, tending continuously to zero. \bigskip

\begin{figure}[h]
\centering\includegraphics[scale=0.3]{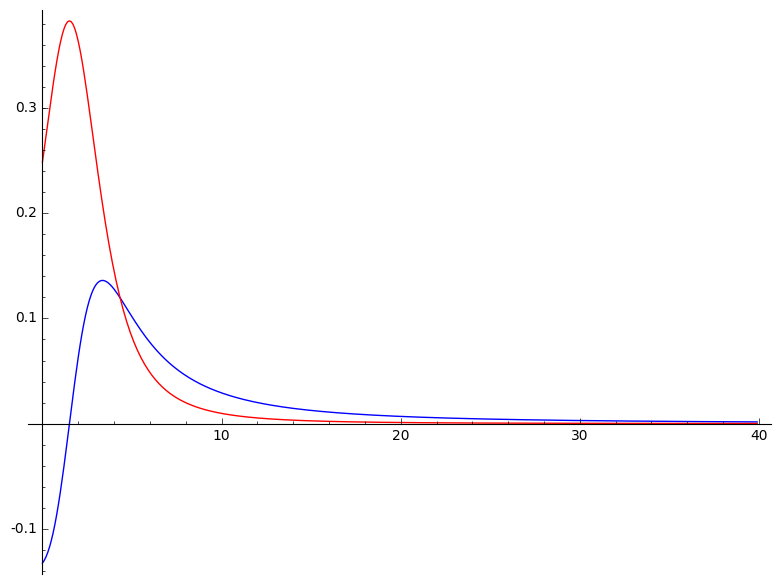} \caption{Graphs of $a_{x}$
$\times$ time (blue), $a_{y}$ $\times$ time (red), corresponding to the
scenario represented in Fig. \ref{fig:case3}.}%
\label{fig:accel3}%
\end{figure}

\textbf{Situation 4:} For $p_{0y}<0$ and $p_{0x}<0$ $\left(  \alpha>0\right)
,$ we notice that $a_{x}(0)<0$, but $a_{x}$ becomes positive from
$t=-p_{0y}/f_{y}$ on; and $a_{y}(t)>0$ for any time. This case is depicted in
Fig. [\ref{fig:accel3}], corresponding to the scenario of Fig. \ref{fig:case3}%
, where $\mathrm{v}_{x}(0)=-0.8c,$ $\mathrm{v}_{y}(0)=-0.5c.$ The product
$\mathrm{v}_{x}\mathrm{v}_{y}$ is initially positive and $a_{x}$ starts
negative. Due to the action of the force $f$, $\mathrm{v}_{x}\mathrm{v}_{y}$
starts to decrease, until $\mathrm{v}_{y}$ and $a_{x}$ become null. From this
point on, $a_{x}>0$, reaches a maximum and then tends continuously to zero.
The graph of $a_{x}$ in Fig. \ref{fig:accel3} is the negative of the one in
Fig. \ref{fig:accel2}, while the acceleration $a_{y}$ displays the same
behavior of the previous case - because it depends only on the absolute value
of velocities. The overall behavior of movement in the x-direction is the
same, acceleration followed by deceleration, with the difference being the
sign of $\mathrm{v}_{x}$ and $a_{x}$. That explains why the graph of $a_{x}$
in Fig. \ref{fig:accel3} is the negative of the one in Fig. \ref{fig:accel2}.

In all of the previously discussed scenarios, the time at which $a_{x}$
achieves its minimum or maximum can be found by differentiating twice eq.
(\ref{eq:vx}) and setting it equal to zero, which yields
\begin{equation}
t=\frac{\sqrt{2m_{0}^{2}c^{2}+2p_{0x}^{2}}-2p_{0y}}{2f_{y}}%
\end{equation}

The acceleration behavior for the case $\boldsymbol{f}=(0,f_{y})$, shown in
Figs. [\ref{fig:accel},\ref{fig:accel4},\ref{fig:accel2},\ref{fig:accel3}],
corresponding to the situations of Figs. [\ref{fig:case1},\ref{fig:case4}%
,\ref{fig:case7},\ref{fig:case3}], reveal a non-monotonic behavior, in
general. By analyzing such graphs, it is correct to claim that the
acceleration absolute value becomes monotonically decreasing after it achieves
a maximum or minimum point, but not since the beginning of movement.

\noindent

\section{Final Remarks}

In this article, we have revisited some aspects of the accelerated motion of a
particle under the action of a constant relativistic force. The asymptotic
velocity in each axis, $(f_{i}/f)c,$ is proportional to the force component,
$f_{i},$ so that after a very long time, force and velocity always tend to be
parallel. In order to fulfill momentum conservation, it is possible to have
acceleration in one axis where there is no force. Previous works in literature
claimed that a particle will exhibit a negative acceleration component in the
direction of the smallest force component, which is not always true. It was
argued that a negative acceleration can exist in the direction of the biggest
force. We have also shown that the acceleration magnitude does not decreases
monotonically, as claimed before. The acceleration can reach a maximum or a
minimum peak before decaying to zero. Several graphs were plotted illustrating
the behavior of velocity and acceleration components for the case of a single
force in the y-axis, $\boldsymbol{f}=(0,f_{y}).$

\end{document}